# Bolometric arrays and infrared sensitivity of VO2 films with varying stoichiometry

Y. Hu,[1] C. H. Lin,[1] S. Min,[1,2] R. L. Smith[2] and S. Roberts[2]

[1]Wuhan Polytechnic University, Wuhan, China, [2]Rio Salado College, Tempe, AZ, USA

## Abstract

Here we propose a linear microbolometric array based on VOx thin films. The linear microbolometric array is fabricated by using micromachining technology, and its thermo-sensitive VOx thin film has excellent infrared response spectrum and TCR characteristics. Nano-scale VOx thin films deposited on SiO2/Si substrates were obtained by e-beam vapor deposition. The VOx films were then annealed at temperatures between 300 to 500 ℃ with various deposition duration time. The crystal structures and microstructures were examined by XRD, SEM and ESCA. These films showed a predominant phase of rhombohedral VO2 and the crystallinity of the VO2 increased as the annealing temperature increased. Integrated with CMOS circuit, an experimentally prototypical monolithic linear microbolometric array is designed and fabricated. The testing results of the experimental linear array show that the responsivity of linear array can approach 18KV/W and is potential for infrared image systems.

## 1. Introduction

In recent years, resistive random access memory (RRAM) devices have more and more attentions because of its advantages, such as low power requirement, fast switching speed, and small size [1-30]. RRAM was made of sandwiched, metal-insulator-metal (MIM), structure. The resistive switching behavior in most of the transition metal oxide (TMO) –based RRAMs was explained by filamentary model [4], where the change of resistance from high resistive state (off-state) to low resistive state (on- state) is called the set process, and the anti-process is called reset process. According to I-V curve of the RRAM device, when set and reset process appear at the same quadrant, the resistive switching behavior was classified as unipolar resistive switch (URS), if not, the resistive switching behavior was classified as bipolar resistive switch (BRS) [31-70]. VO2 has been studied several years, since F. J. Morin found its semiconductor-to-metal transition temperature Tt = 340 K in 1959. This transition property makes VO2 good applications for encompassing thermochromic coatings, optical and holographic storage systems, fiber optical switching devices, and laser scanners, but not for RRAMs. [71-140]

Microbolometric arrays have more advantages such as higher stability, reduced power dissipation, smaller size and lighter weight than conventional cooled infrared detector arrays. Development work on microbolometric arrays has focused on improving their sensitivity. One of key methods for improving microbolometric sensitivity is through mending the process of fabricating microbolometers to increase their temperature coefficient of resistance (TCR) values. Since the phase-transition phenomenon of vanadium dioxides (VO2) materials was first observed in the late 1950s, many researchers have been interested in VO2 thin films. Early tests of TCR indicated that the films of VO2 would perform better than a metal TCR resistor, furthermore the TCR values of VO2 thin films can be controlled by adjusting their ingredients. In addition, thin films of VO2 are selected for microbolometeric arrays because the fabricating technology of VO2 thin films is compatible with integrated circuit (IC) process, which is favorable to designing and fabricating monolithic microbolometric arrays.

Table 1: The process flow details

| Target | $V_2O_5$ powder |
|---|---|
| Pre-evaporating Time | 3 min |
| Evaporating Time | 20 sec |
| Emission | 53 % |
| Pressure | $5 \times 10^{-5}$ Torr |
| Substrate Temperature | 32 °C |

The performance of VO2-based microbolometeric arrays depends on not only the TCR of VO2 thin films but also the structure of microbolometers and the performance of the readout integrated circuit. For linear microbolometric array, the simple structure of conventional micro-bridge is unfavorable to microbolometric sensitivity. The conventional micro-bridge structure for linear microbolometric arrays is shown in Fig. 1, in which the thermal conductance of the supporting legs is fairly large and can lower the microbolometric sensitivity. Besides, it is also a challenge to achieve good performance of the readout integrated circuit. In this paper, a new linear microbolometric array based on vanadium oxides (VOx) thin film is designed and fabricated. Integrated with complementary metal oxide semiconductor (CMOS) readout integrated circuit, the VOx microbolometers in linear array has more perfect structure compared with conventional micro-bridge structure for linear array and excellent TCR characteristics. The experimental monolithic linear arrays based on VOx thin film reveal good performance and can be applied to practical infrared image systems.

## 2. Experimental details and Fabrication

We used Si (100) wafer which was oxidized in a furnace to grow a SiO2 insulator layer as the substrate. Bottom electrodes of Pt were patterned with a shadow mask by sputter. The evaporating target was made of V2O5 powder, and then we deposited VOx thin films on Pt/SiO2/Si substrate under 5x10-5 torr for 20 second by evaporator. The coating conditions were shown in Table 1. After evaporating VOx thin films, we used Rapid Thermal Annealing (RTA) to anneal VOx thin films under 10 mtorr at different temperatures (300/400/450/500 °C) for 1 minute. The top Pt electrodes were deposited on VOx thin films at the end. The crystalline structure of VOx thin films were determined by X-ray diffraction (XRD). The binding energy of V2O3 thin films were measured by Electron Spectroscopy for Chemical Analysis (ESCA). The cross-sectional morphology of the V2O3/SiO2/Si structure was obtained using a Field Emission Scanning Electron Microscopy (FESEM). The resistive switching behaviors of metal-insulator-metal (MIM) devices were measured by Keitheley-4200.

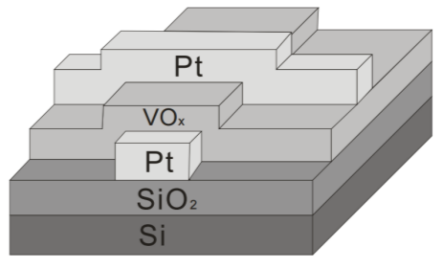

Fig. 1: Schematic of the device structure described in the text.

In order to improve the sensitivity of microbolometer, a standard micro-bridge structure which is widely utilized in microbolometric focal plane arrays is used for microbolometer pixel [6, 7]. The micro-bridge structure of VOx bolometer is illustrated in Fig. 2. The microbolometer consists of a 0.5 μm thick micro-bridge of Si3N4 suspended above the underlying silicon substrate, two narrow legs of Si3N4 which are 2.0 μm high and supporting the bridge, and a 100×100 μm thin film (0.5 μm thick) of VOx which is encapsulated in the center of the micro-bridge. The basic operational principal of the microbolometer is as follows. When the VOx thin film is irradiated by infrared (IR) radiation and heated, its resistance value is changed. This change in the resistance value of the VOx thin film can be converted to voltage signal to be output through the CMOS readout integrated circuit in silicon substrate, which is connecting to the narrow legs via the X or Y metal on the surface of silicon substrate. From Fig. 2, it can be found that the two supporting legs are narrow and clival.

This kind of supporting legs is in favor of the microbolometer heat isolation from silicon substrate, so the IR heating efficiency on the VOx thin film is high, which can greatly improve microbolometric sensitivity.

The micro-bridge structure microbolometer is produced in a way using a surface micromachining technique. Figure 3 illustrates the six fabrication steps of a micro-bridge structure microbolometer. Fabrication begins with implantation of the required CMOS readout electronics and conducting metallizations in the silicon wafer. The wafer is then planarized with the spun-on polyimide, which is photolithographically patterned to form sacrificial mesa. Over the sacrificial mesa the silicon nitride layer is sputtered together with subsequent VOx thin film. Subsequently, together with connecting metallizations the silicon nitride protecting layer is sputtered. And as a final step, the sacrificial mesa is removed by oxygen plasma etching to leave a self-supporting micro-bridge structure.

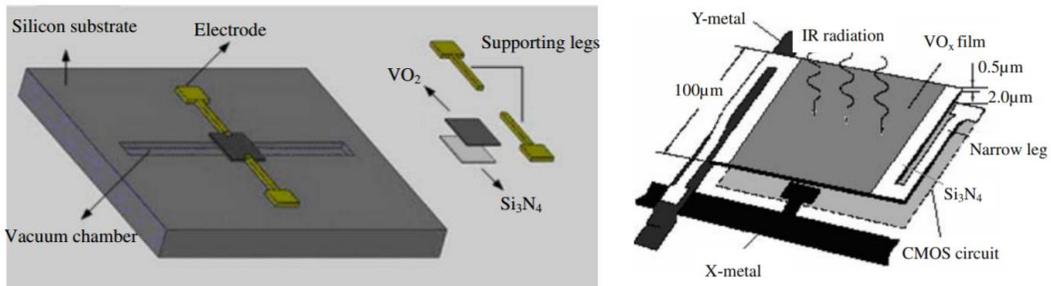

Fig. 2: (left) Schematic of the device fabrication process and (right) magnified view of the crab-leg structure.

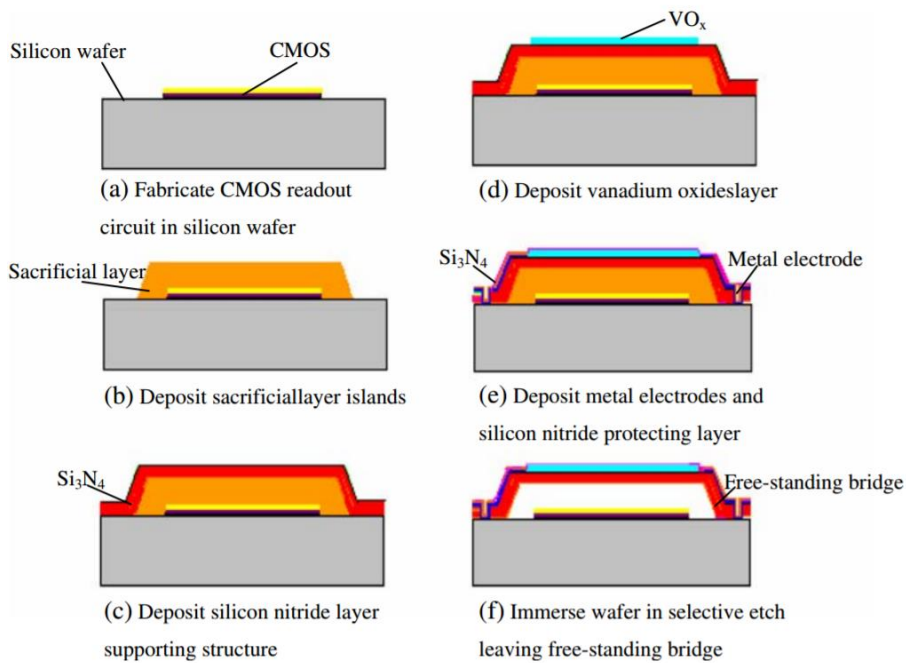

Fig. 3: Illustration of the process flow.

## 3. Results and discussions

The micro-bridge structure design allows great fill factor for the CMOS readout integrated circuit is placed in the silicon underneath the micro-bridge and great IR absorption for the underlying cavity produces a resonant optical cavity. So this type of micro-bridge structure design can improve the whole performance of linear array including increasing microbolometric sensitivity and decreasing noise equivalent temperature difference, that is to say that little IR radiation power from target can produce great resistance change of the VOx thin film. Among the components of micro-bridge structure microbolometer, the photoelectric properties of VOx thin film are vital, and the VOx thin film fabricated in our laboratory has excellent optical and electrical properties. On the fourth step in the fabricating processes of microbolometer, a thin film of vanadium produced at first undergoes a chemical reaction by being annealed, which leads to form the VOx thin film. In fact, the annealing is in Ar and O2, so the formed thin film consists of VOx consequentially. Because the relationship curve of temperature-resistance of the thin film is different from the relationship curve of VO2 material [3], the thin film does not consist of VO2 material only. X-ray diffraction (XRD) measurement makes it clear that the thin films contain V2O3 except for VO2 rather than VO2 only. This is showed in Fig. 4, from which it is found that the VOx thin film is the mixture of VO2 and V2O3. Experiments demonstrate that the different annealing temperature and time can result in the different amount of V2O3 material contained in the VO2 thin film, so the fabricating process of VOx is essentially a process which is adjusting the ingredients of the thin film. The long wave infrared spectrum absorptivity of the VOx thin film, in the 8.0–15.0 μm wavelength range, is shown in Fig. 5.

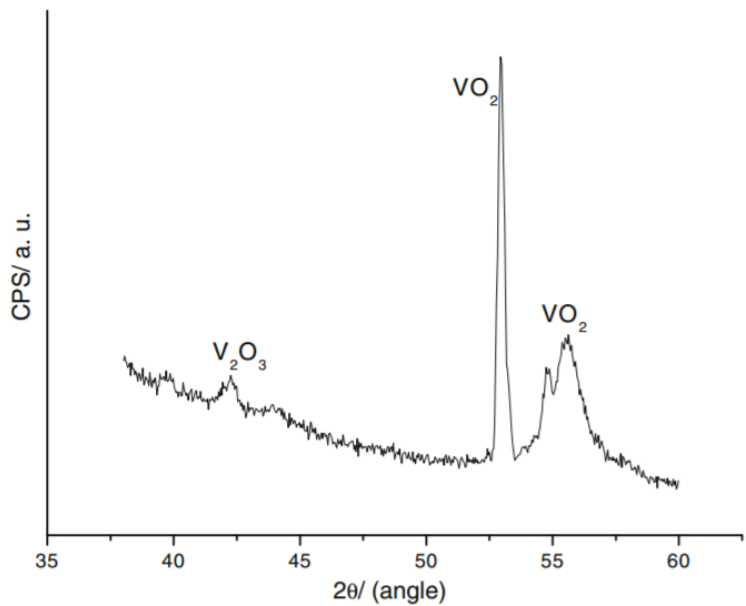

Fig. 4: XRD spectrum of the VO2 film.

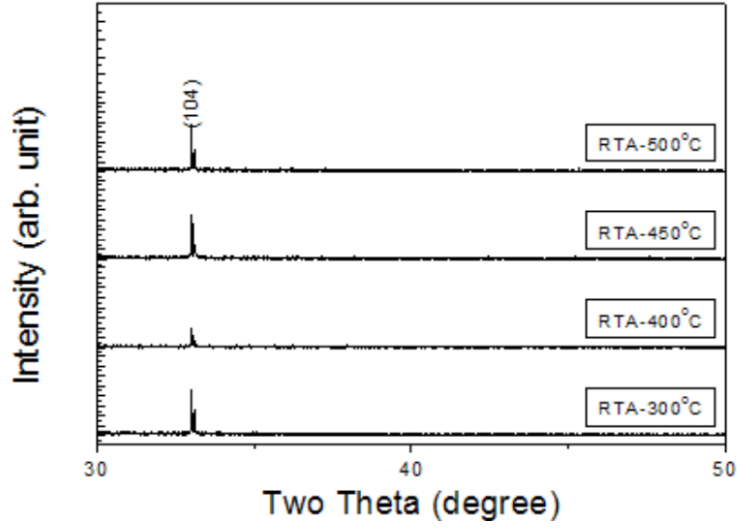

Fig. 5: XRD of the VO2 films at different RTA temperatures used.

It can be found that the VOx thin film has very high infrared absorptivity in long wave infrared spectrum. At the same time, measurement shows that the VOx thin film has excellent temperature-resistance characteristic. Figure 6 shows the curve of sheet resistance versus temperature for the VOx thin film, from which it can be found that the TCR of the VOx thin film reaches to −2.7% at room temperature and is higher than that of traditional VO2 thin film (about −2.0%) [2]. In addition, the CMOS readout integrated circuit of the linear array adopts a combined readout structure of bolometer current direct injection (BCDI) and capacitor feedback transimpedance amplifier (CTIA). The CMOS readout integrated circuit is shown in Fig. 7. The BCDI input circuit is formed by the bolometer Rd (when exposed to infrared light, its resistance can be changed), the blind bolometer Rb (not exposed to infrared light), the PMOS device P1 and the NMOS devices N1. The CTIA integrated circuit contains the reset switch N2, the integration capacitor Cint, and the amplifier A. The blind bolometer Rb can cancel the bias current contained in the bolometer Rd and depress the affection of substrate temperature fluctuation, the MOS devices N1 and P1 are a pair of complementary integration controlling switches, which are utilized to control the integration time of the photon-generated current Iint, and the CTIA provides a highly stable bias to the detector and has high-performances such as high-gain, high photon current injection efficiency and low noise. The circuit operation is explained as follows. When the pair of integration controlling switches N1 and P1 are ON and the reset switch N2 is OFF, the photon-generated current Iint is integrated into the integration capacitor Cint. After an integrating interval, the pair of integration controlling switches N1 and P1 turn off and the readout voltage signal Vout is output.

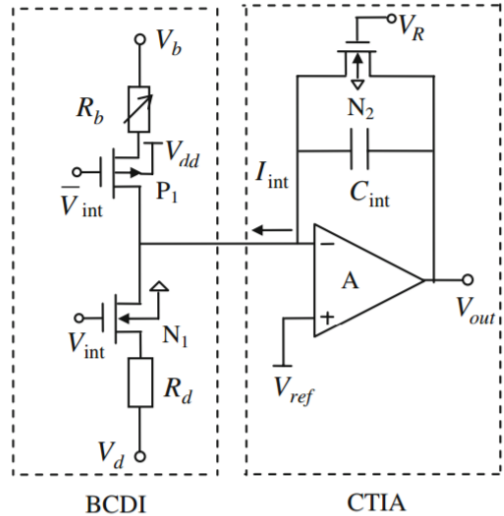

Fig. 6: Circuit schematic of the bolometer.

An experimental monolithic chip for 100×100 μm-pitch 128 element linear microbolometric array has been designed and fabricated to verify the designed linear array based on VOx thin film. The micrograph of two typical microbolometers in the linear array is shown in Fig. 8. This linear array is designed to work under 5 V power supplies, packaged in vacuum and tested under a 298 K environment. The voltage responsivity of the linear array is defined in Equation 1 where d is the number of bad and dead detectors, and Rv(I) is the voltage responsivity of the microbolometer No. I and expressed as RvðIÞ ¼ VoutðIÞ P in which Vout(I) is the output voltage of the microbolometer No. I heated by infrared irradiation power P. Measurements show that the zero-frequency voltage responsivity of the linear array is the function of bias current in microbolometer which is illustrated in Fig. 9. When the bias current increases above 20 μA, the zero-frequency voltage responsivity is almost constant (approximately about 18KV/W), but being accompanied with gradually increasing noise. By maintaining 20 μA bias current, the voltage responsivity at different frequencies is obtained. The relationship between voltage responsivity and frequency is shown Fig. 10, from which it can be found that the voltage responsivity declines with the increase of signal frequency. The Nyquist frequency of the voltage responsivity is about 10 Hz, which is consistent with designed response time characteristic.

Fig. 1 shows a part of metal- insulator-metal (MIM) structure. Fig. 2 shows that the XRD patterns of the VOx thin films that they were deposited on Pt/SiO2/Si substrate annealed at different temperature (300/400/450/500 °C) by RTA. All of thin films showed the same orientation peak at 2θ=33.0o which were matched with JCPDS card No. 85-1411. Basing on the result that all of thin films were V2O3 with rhombohedral structure. Fig. 3 is the ESCA patterns of the VOx thin films which were annealed at 300/400/450/500 °C by RTA.

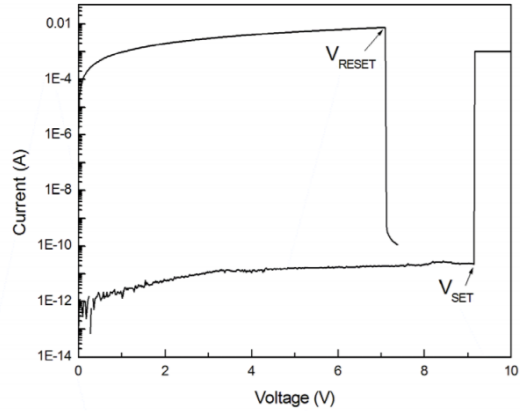

Fig. 7: Current-voltage curves showing the set and reset process

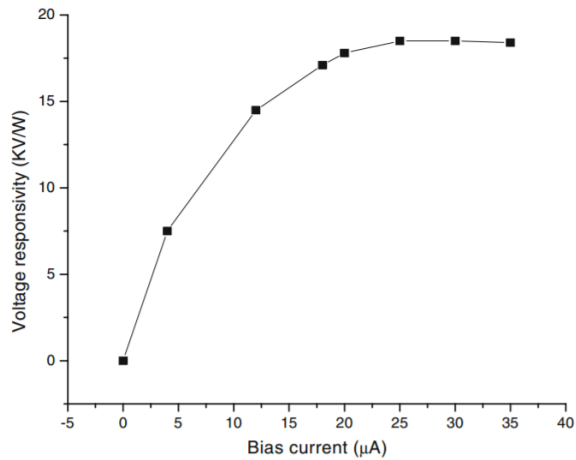

Fig. 8: Voltage responsivity against bias current.

The peaks at 516.3 eV are the binding energy of V2O3 where the valence state of V is V+1.5. The SEM of the V2O3 surface morphology annealed at 400 o C was shown in Fig. 4. The surface of V2O3 thin film was smooth, and there do not appear obviously crystal partical. Fig. 5 shows the cross section of V2O3 films annealed at (a) 300 (b) 400 (c) 450 (d) 500 ℃ by RTA. The thickness of V2O3 thin films were about 150~200 nm, and there had fine V2O3/SiO2 interface. Fig. 6 shows the resistive switching characteristics of Pt/ V2O3/Pt/SiO2/Si device where V2O3 thin films were annealed at 400 ℃ for 1 minute by RTA. It show that the Imax (Vreset) is 7.93x10-3 A, and Iset (Vset) is 2.26x10-11 A, the current range between Vreset to Vset is about 7 order. Summary We observed the thin film with rhombohedral structure after annealing at different temperatures by RTA. The thickness of the thin films is about 150～200 nm depended on the deposition duration times. In this work, we found the device with Pt/V2O3/Pt structures annealed at 400 ℃ exhibit the largest memory window. The range of the memory window (ION to IOFF) achieved nine orders. The mechanism of the resistive switching behavior of the V2O3 thin film was explained by the filamentary model.

## 4. Conclusions

A linear microbolometric array based on VOx thin film has been fabricated by using a new micromachining process, which consists of six steps. The benefit of the new micromachining process is to improve the microbolometric sensitivity and this leads to high voltage responsivity of the linear microbolometric array. We expect the linear micro-bolometric array based on VOx thin film will play an important role in developing potential civil and military applications. Further investigations on improving the performance characteristics such as noise equivalent temperature difference and nonuniformity are in progress.